\documentclass{PoS}

\usepackage{graphicx}
\usepackage{amsmath,amssymb,amsfonts}
\usepackage{bbm}
\usepackage{bm}
\let\normalcolor\relax

\newcommand{\mlatt}{m_\text{latt}}
\newcommand{\ev}[1]{\left\langle #1 \right\rangle}
\newcommand{\bbZ}{\mathbb{Z}}
\newcommand{\Ns}{{N_\mathrm{s}}}

\newcommand{\ord}{\mathcal{O}}
\newcommand{\Ncal}{\mathcal{N}}
\newcommand{\half}{{\textstyle\frac{1}{2}}}
\newcommand{\SB}{S_\text{B}}

\newcommand{\myindent}{\noindent}

\title{Numerical Investigation of the 2D $\pmb{\mathcal{N}}\mathbf{\!=\!2}$ Wess-Zumino Model}

\ShortTitle{Numerical Investigation of the 2D $\mathcal{N}\!=\!2$ Wess-Zumino Model}

\author{\speaker{Christian Wozar}, Georg~Bergner, Tobias~K\"astner,
Sebastian~Uhlmann, Andreas~Wipf\\
        Theoretisch-Physikalisches Institut,
Friedrich-Schiller-Universit{\"a}t Jena, Max-Wien-Platz 1, 07743
Jena, Germany\\
        E-mail: \email{christian.wozar@uni-jena.de}}

\abstract{We study lattice formulations of the two-dimensional $\mathcal{N}=2$ Wess-Zumino
model with a cubic superpotential. Discretizations
with and without lattice supersymmetries are compared. We observe
that the ``Nicolai improvement'' introduces new problems to simulations of the
supersymmetric model. \newline
With high statistics we check the degeneracy of bosonic
and fermionic masses on the lattice. 
Perturbative mass corrections to
one-loop order are compared with continuum extrapolations of our lattice results in the
weakly coupled regime.
For intermediate couplings
first results of fermionic masses in the continuum are presented where 
deviations from the
perturbative result are observed.}

\FullConference{The XXVI International Symposium on Lattice Field Theory\\
                 July 14-19 2008\\
                 Williamsburg, Virginia, USA}

\begin{document}


\section{Introduction}
\myindent
\setlength{\baselineskip}{0.97\baselineskip}%
The two-dimensional $\Ncal=2$ Wess-Zumino model in the continuum shows no
spontaneous supersymmetry breaking. However, a lattice formulation must break
(part of) the
supersymmetry explicitly due to the failure of the Leibniz rule on the lattice. In
this model high statistics
on large lattices are available and supersymmetry restoration effects can be
analyzed numerically. The restoration of supersymmetry in the continuum limit
must be treated carefully as has been demonstrated
in supersymmetric quantum mechanics \cite{Bergner:2007pu,Kastner:2007gz}. One
possible way to circumvent
relevant supersymmetry breaking operators is the application of a blocking
transformation to a free theory \cite{Bietenholz:1998qq} leading to solutions
similar to the
Ginsparg-Wilson relation for the chiral symmetry \cite{Bergner:2008ws}.

A further suggestion keeps at least one supersymmetry on the lattice preserved and
goes under the name of
``Nicolai improvement'' \cite{Nicolai:1979nr}. In former works
\cite{Beccaria:1998vi,Catterall:2001fr} 
such improved models using Wilson fermions were simulated, and
discrepancies to the perturbative result as well as problems with the
extraction of masses occurred at stronger couplings.
In this work (see \cite{Kastner:2008zc} for detailed analyses) the effects of
the Nicolai improvement in the intermediate coupling regime are analyzed and
compared to results of \emph{unimproved} simulations. Additionally different
fermion formulations (standard/twisted Wilson, SLAC) are explored.


\section{The model}
\begin{floatingfigure}[r]
\vspace{-1ex}
\includegraphics{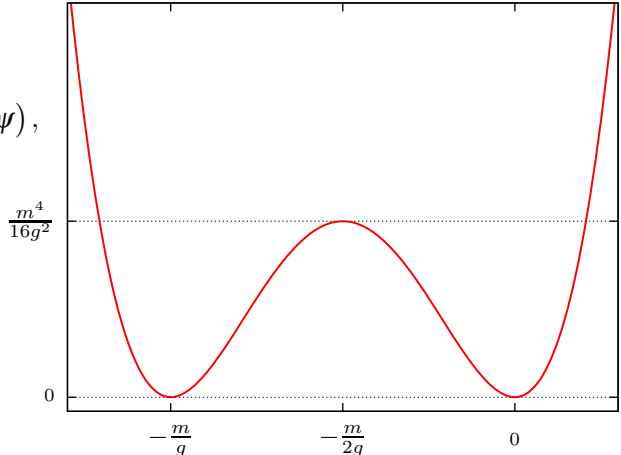}
\caption{\label{fig:classPot}Classical potential $|W'(\varphi_1)|^2$  from
  \eqref{eq:superpot} for vanishing imaginary part
  ($\varphi_2=0$). In the free theory limit ($g\to 0$) the
  left minimum is pushed towards minus infinity.}
\end{floatingfigure}
\myindent
The continuum action with complex field
$\varphi=\varphi_1+i\varphi_2$,
\begin{equation}\label{eq:superpot}
\begin{aligned}
S_\text{cont} &= \int d^2x
\left(
2\bar\partial\bar\varphi\partial\varphi+
\half |W'(\varphi)|^2 +\bar\psi M\psi
\right),\\ M&=\gamma^z\partial + \gamma^{\bar z}\bar\partial + W''P_{+}
+\overline{W}''P_{-}
\end{aligned}
\end{equation}
is invariant
under \emph{four real supercharges}. Taken together they satisfy the 
$\mathcal{N}=(2,2)$ superalgebra, and it has been argued
that one supersymmetry can be preserved on the lattice \cite{Catterall:2001fr}.

The holomorphic superpotential here (see Fig.~\ref{fig:classPot})
\begin{equation}
W(\varphi) =
\half m\varphi^2+\textstyle{\frac{1}{3}}g\varphi^3
\end{equation}
contains a mass parameter $m$ and defines a dimensionless coupling
$\lambda=\frac{g}{m}$. This theory possesses a discrete $\bbZ_2^4$ symmetry
which in general is partially broken by a lattice discretization. 
A perturbative expansion in orders of $\lambda$ around the free theory at $\lambda=0$ is
possible and is used below.


\subsection{Nicolai improvement}
\myindent
\emph{One real supersymmetry} can be preserved on the lattice by using the action
\begin{equation}
S = \half\sum_x \bar\xi_x\xi_x + \sum_{xy}\bar\psi_x M_{xy}\psi_y
\end{equation}
in terms of the Nicolai variable $\xi_x=2(\bar\partial\bar \varphi)_x+W_x$
with $W_x=W'(\varphi_x)$, $W_{xy}:=\partial W_x / \partial\varphi_y$ and
\begin{equation}
M_{xy}=\begin{pmatrix}
         W_{xy} & 2\bar\partial_{xy}\\
         2\partial_{xy} & \overline{W}_{xy} 
         \end{pmatrix}
= \renewcommand{\arraystretch}{1.4} \begin{pmatrix}
         \frac{\partial \xi_x}{\partial\varphi_y}&  \frac{\partial
         \xi_x}{\partial\bar\varphi_y} \\
         \frac{\partial \bar\xi_x}{\partial\varphi_y}&  \frac{\partial \bar\xi_x}{\partial\bar\varphi_y}  
         \end{pmatrix}.
\end{equation}
In terms of the original field $\varphi$, the lattice action reads
\begin{equation}
S=\sum_x \Big(2 \left(\bar\partial\bar\varphi\right)_x(\partial\varphi)_x +
\half\big|W_x\big|^2 + W_x(\partial\varphi)_x +
\overline{W}_{\!x}(\bar\partial\bar\varphi)_x\!\Big)+\sum_{xy}\bar\psi_x
M_{xy}\psi_y.
\end{equation}
This action only differs from a straightforward
discretization by discretized \emph{surface terms}
\begin{equation}
\Delta S = \sum_x \Big(W_x(\partial\varphi)_x+
\overline{W}_{\!x}(\bar\partial\bar\varphi)_x\Big)
\end{equation}
which must vanish in the continuum limit.


\subsection{The lattice discretization}
\myindent
To preserve the full supersymmetry of the free theory the
same lattice derivatives for
bosonic and fermionic degrees of freedom must be used. In this extensive study we
compare three different lattice derivatives:
\begin{itemize}
  \item Symmetric derivative
  $\left(\partial^\text{S}_{\mu}\right)_{xy}=\half(\delta_{x+\hat\mu,y}-\delta_{x-\hat\mu,y})$ with
  \emph{standard Wilson} term $W_x = W'(\varphi_x) - \frac{r}{2}
  (\Delta\varphi)_x$ using $r=1$. The Wilson term must be added to
  $W_x$ (and not to the derivative) in order to obtain an antisymmetric matrix $(\partial_\mu)_{xy}$.
  This results in a fermion matrix
  \begin{equation}
  M_{xy} = \begin{pmatrix}
          W''(\varphi_x)\delta_{xy} & 2\bar\partial_{xy}\\
         2\partial_{xy} & \overline{W}''(\bar\varphi_x) \delta_{xy}
         \end{pmatrix} -
\frac{r}{2}\Delta_{xy} .\end{equation}
  \item Symmetric derivative $\partial^\text{S}$ with \emph{twisted (imaginary)
  Wilson} term $W_x = W'(\varphi_x) + \frac{i r}{2} (\Delta\varphi)_x$ resulting
  in
  \begin{equation}
  M_{xy}= \begin{pmatrix}
         W''(\varphi_x)\delta_{xy} & 2\bar\partial_{xy}\\
         2\partial_{xy} & \overline{W}''(\bar\varphi_x) \delta_{xy}
         \end{pmatrix} + \gamma_3\frac{r}{2}\Delta_{xy} .\end{equation}
 The choice $r=2/\sqrt{3}$ reproduces the mass of the free theory up to 
 $\ord(a^4)$ at lattice spacing $a$ as discussed in \cite{Bergner:2007pu}.
  \item \emph{SLAC derivative} $\partial_{x\neq
y}=(-1)^{x-y}\frac{\pi/N}{\sin(\pi(x-y)/N)}$, $\partial_{xx}=0$ with 
fermion matrix
  \begin{equation}
  M_{xy}= \begin{pmatrix}
         W''(\varphi_x)\delta_{xy} & 2\bar\partial_{xy}\\
         2\partial_{xy} & \overline{W}''(\bar\varphi_x) \delta_{xy}
         \end{pmatrix}.
 \end{equation}
\end{itemize}
Using these discretizations, we have simulated the improved and
unimproved (without surface terms) models 
applying a \emph{combination of Fourier accelerated HMC with higher-order
integrators}.


\section{Limitations of the Nicolai improvement}
\begin{figure}
\includegraphics[width=0.47\textwidth]{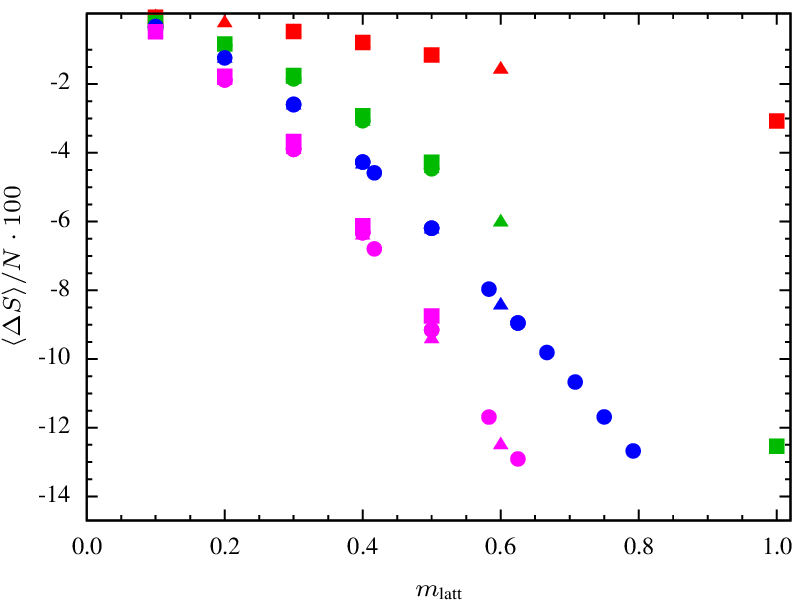}\hfill
\includegraphics[width=0.47\textwidth]{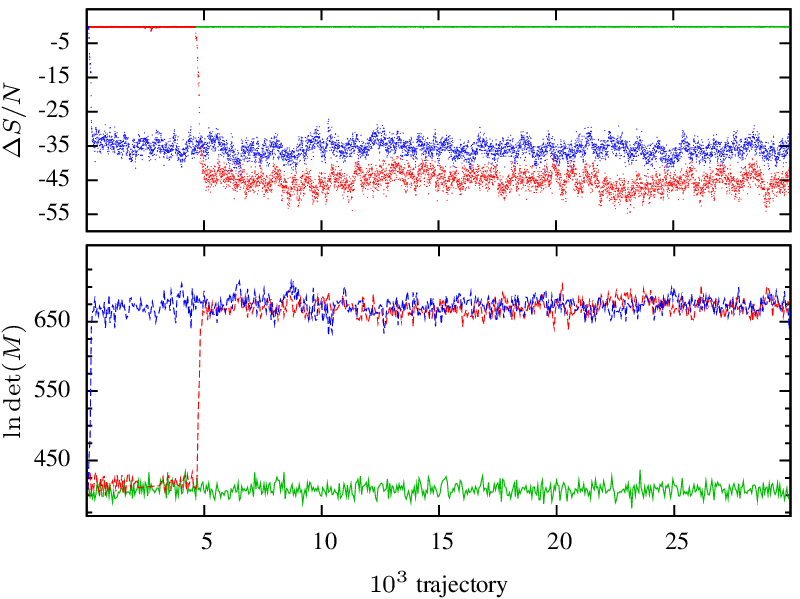}
\caption{\label{fig:slac_improvement}\textsl{Left:} Reduced improvement term
$\Delta S/N$ for different lattice sizes: $9\times 9$
  (squares), $15\times 15$ (triangles) and $25\times 25$ (circles). 
  Colors depict $\lambda=$ $0.8$ (red), $1.0$ (green), $1.2$
  (blue),  $1.5$ (magenta).\newline \textsl{Right:} MC history of improvement term and
  fermion determinant (SLAC improved, $N=15\times
  15$, $\mlatt=0.6$, $\lambda=$ $1.4$ (green), $1.7$ (red),
  $1.9$ (blue)).}
\end{figure}
\begin{floatingfigure}[r]\vspace{-1.5ex}
\includegraphics{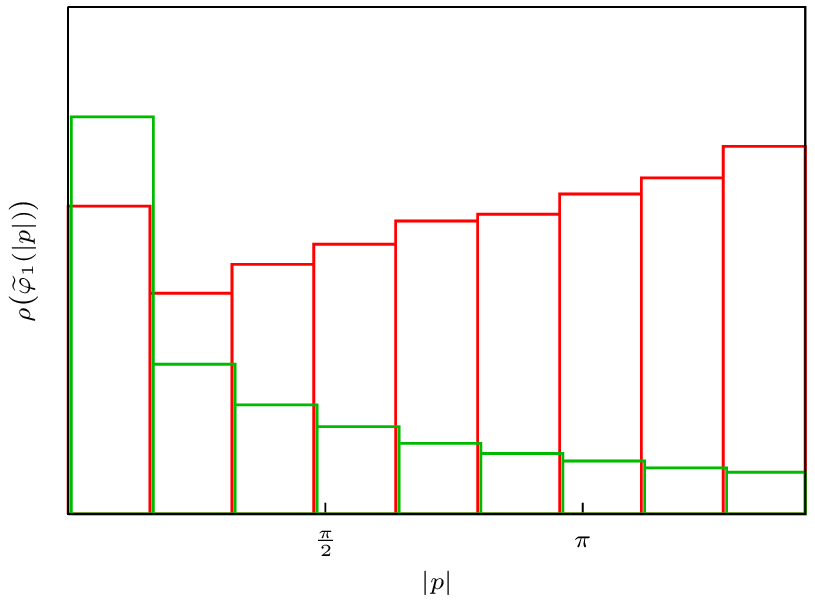}
\caption{\label{fig:mode_analysis}Mode analysis of ensembles in the
  physical (green, $\lambda=1.4$) and unphysical (red, 
  $\lambda=1.7$) phase. Here $\rho$ is the distribution function
  for the modulus of the lattice momentum averaged over 25,000 configurations
   (SLAC improved, $N=15\times 15$, $\mlatt=0.6$).}
\end{floatingfigure}
\myindent
For simulations of the improved model including dynamical fermions the
expectation value of the bosonic action is independently of $\lambda$ fixed to
\begin{equation}
\ev{\SB} = N = \text{\# lattice points}.
\end{equation}
Nevertheless the improvement term $\Delta S = \sum_x
\Big(W_x(\partial\varphi)_x+ \overline{W}_x(\bar\partial\bar\varphi)_x\Big)$
does not necessarily vanish. Therefore, we analyze $\Delta S$ with SLAC fermions at
different couplings and for different lattice masses $\mlatt$ $=m/\Ns$
(Fig.~\ref{fig:slac_improvement}, left panel). Here
$\Ns$ denotes the number of lattice points in spatial direction. In the continuum
limit ($\mlatt$ $\rightarrow 0$) the improvement term consistently vanishes for
every $\lambda$.

For $\ev{\Delta S}/N > 14\%$ the behavior of improvement
term and fermion
determinant changes significantly (Fig.~\ref{fig:slac_improvement}, right
panel).
The improvement terms dominates the bosonic action by more than one order of
magnitude while $\ev{\SB}=N$ is still preserved. Additionally the fermion determinant
grows
drastically and so hinders the system from returning into the original region of
configuration space. This instability can be explained by reconsidering the improved action 
\begin{equation}
\label{eq:impr2}
\SB=\frac{1}{2}\sum_x \Big| 2(\partial\varphi)_x +
\overline{W}_{\!x} \Big|^2 = \sum_x \Big(2
\left(\bar\partial\bar\varphi\right)_x(\partial\varphi)_x +
\half\big|W_x\big|^2\Big) +\Delta S.
\end{equation}
This action allows for two distinct behaviors
of the fluctuating fields. The physically expected behavior consists
of small fluctuations around the classical minima of the potential.
Alternatively, \eqref{eq:impr2} allows for large fluctuations of
kinetic and potential term to be compensated by the improvement term of opposite
sign. In this
situation, it is definitely no longer possible to extract meaningful physics.

Analyzing the distribution of the fields in momentum space
at $\lambda=1.4$ and $\lambda=1.7$ (Fig.~\ref{fig:mode_analysis}) shows that for
too large couplings $\lambda$ (or lattice
masses $\mlatt$) the simulation samples only \emph{unphysical UV dominated}
configurations. Therefore at strong
 coupling a \emph{careful analysis of the improvement
term} during the simulation must be ensured in order to achieve reasonable
simulations.


\section{Weak coupling}
\myindent
In the regime of weak couplings ($\lambda\le 0.4$) we are able to match bosonic
and fermionic masses so as to observe how well supersymmetry effects (e.g.\ the
degeneracy of masses) are realized on the lattice. Furthermore continuum
extrapolations of the different discretizations are compared to the result of
continuum perturbation theory at one-loop order.


\subsection{Signs of supersymmetry at finite lattice spacing}
\begin{figure}
\includegraphics[width=0.47\textwidth]{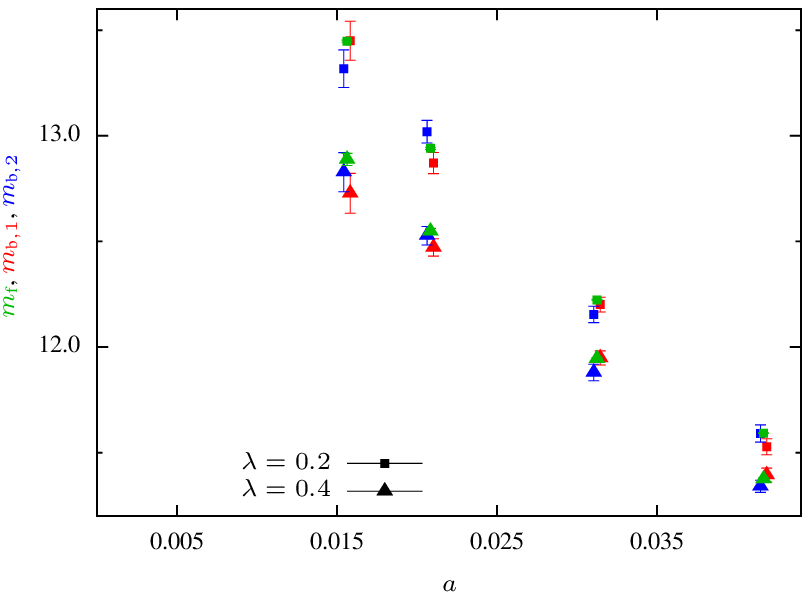}\hfill
\includegraphics[width=0.47\textwidth]{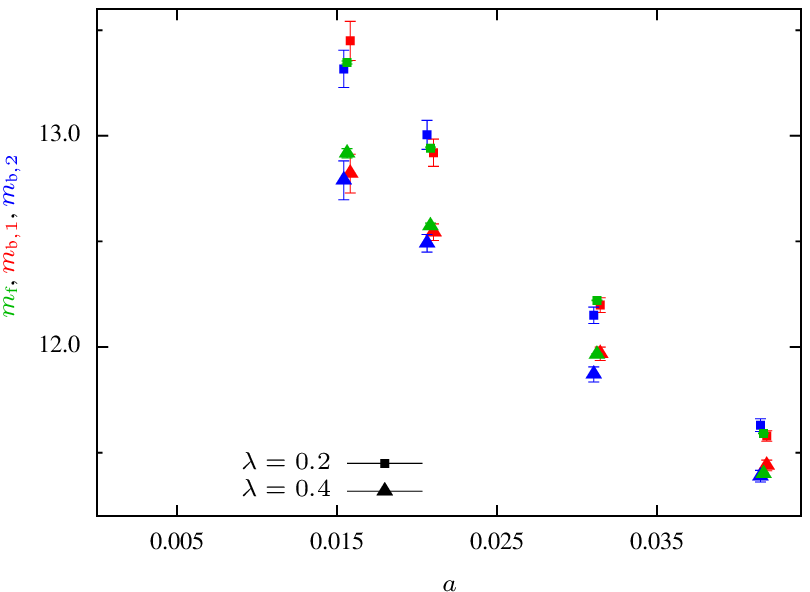}
\caption{\label{fig:bosonFermion}Masses for bosons ($\varphi_1$, $\varphi_2$, statistics
$10^6$--$10^7$ configs)
and fermions (statistics $10^4$ configs) for improved \textsl{(left)} and
unimproved \textsl{(right)} model with standard Wilson fermions.}
\end{figure}
\myindent
In an unbroken supersymmetric theory bosonic and fermionic masses coincide. In
the lattice formulation the supersymmetry is broken explicitly (at least partially).
This induces a possible breaking of the mass multiplets which is
explored at
different lattice spacings for $\lambda\in\{0.2,0.4\}$, $m=15$ (Fig.~\ref{fig:bosonFermion}).
Even with a statistics of up to $10^7$ configurations the masses
of bosons and fermions can
\emph{not be distinguished} in our simulations. Additionally improved and unimproved
models give the same results (within error bars) for $\lambda\le 0.4$.

This demonstrates that for Wilson type fermions in a region where the simulations do
not show unphysical UV effects the Nicolai improvement is \emph{not necessary}.
A stable simulation with the unimproved model is likely to provide the same
results, at least in the continuum limit.


\subsection{Continuum extrapolation}
\begin{figure}
\includegraphics[width=0.47\textwidth]{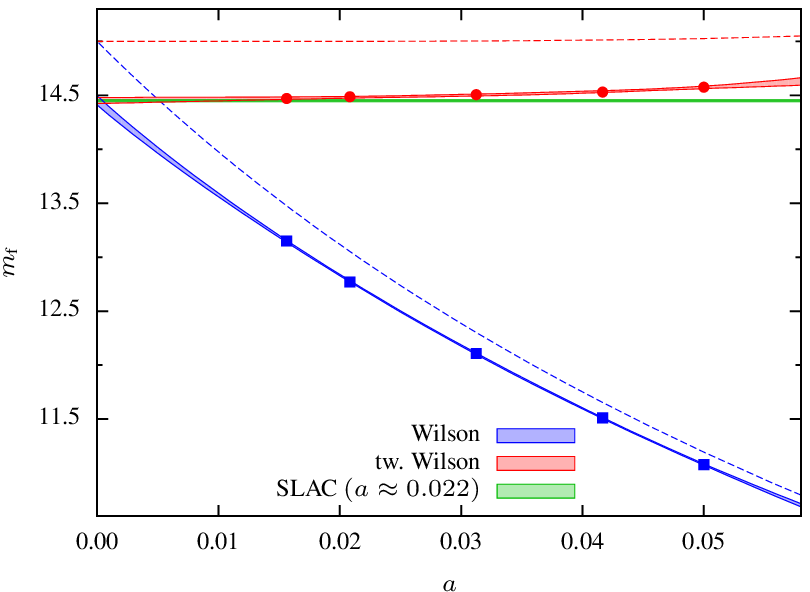}\hfill
\includegraphics[width=0.47\textwidth]{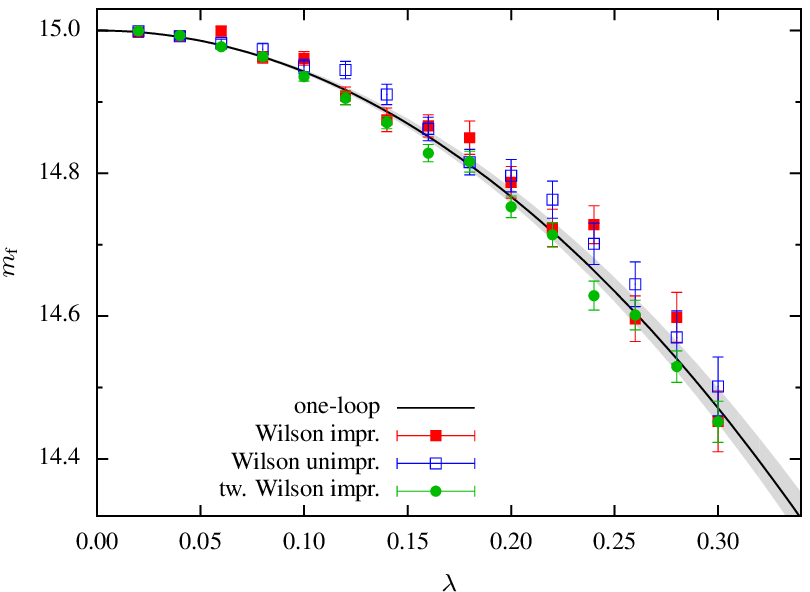}
\caption{\label{fig:interactingContinuum} \textsl{Left:} The continuum
extrapolation of fermionic masses for $\lambda=0.3$ for the improved Wilson
and twisted Wilson model. Here, the SLAC result is given for one single lattice size.
For comparison the exact results for the free theory are also shown.\newline
\textsl{Right:} Continuum extrapolation of fermionic masses for the
weakly coupled regime in comparison to the perturbative result.}
\end{figure}

\myindent
For the free theory the lattice masses can be computed analytically. To make contact
with perturbation theory which is carried out in the continuum it is crucial to
get a stable continuum extrapolation even for the interacting case. 
Extrapolations from finite lattice spacing to the continuum using standard
and twisted Wilson fermions for the improved model ($m=15$, $\lambda=0.3$) are
shown in Fig.~\ref{fig:interactingContinuum} (left panel). These are based on lattice
sizes $\Ns\in\{20,24,32,48,64\}$ and demonstrate that both formulations yield the same
continuum result. Additionally this result also coincides with a prediction by
the SLAC model on a finite lattice ($N=45\times 45$).
Furthermore the twisted Wilson fermions are much closer to the continuum limit
than standard Wilson fermions
for finite lattice spacing. Therefore our analysis of the intermediate coupling case
uses only the twisted type of Wilson fermions and SLAC fermions.

\begin{floatingfigure}[r]
\includegraphics{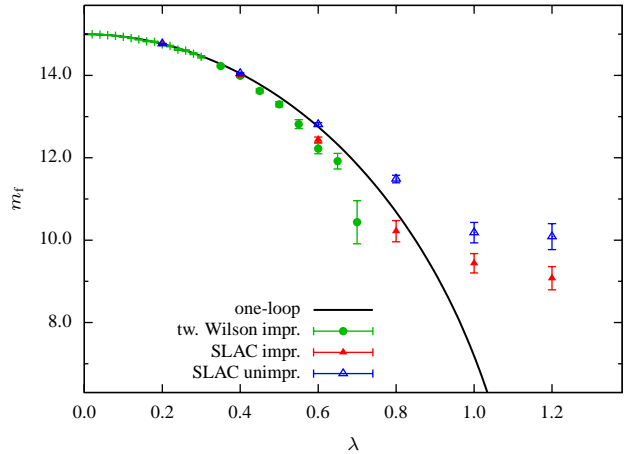}
\caption{\label{fig:slacStrong}Masses of the improved and unimproved model
  with SLAC fermions on a $45\times 45$ lattice and continuum extrapolated results
  for
  twisted Wilson fermions are compared with the perturbative one-loop result in
  the continuum.}
\end{floatingfigure}


\subsection{\mbox{Comparison to perturbation theory}}
\myindent
For small $\lambda$ we compare the
perturbative one-loop result for the renormalized mass
\begin{equation}
m_\text{ren}^2  = m^2
\left(1-\frac{4\lambda^2}{3\sqrt{3}}\right) +\ord(\lambda^4)
\end{equation}
to the continuum extrapolation of the lattice data
(Fig.~\ref{fig:interactingContinuum}, right panel). 
All different formulations are seen to \emph{coincide with perturbation
theory}. Even for the unimproved model with Wilson fermions the correct
(supersymmetric) continuum limit is reached within error bars.


\section{Intermediate coupling}
\myindent
To explore the limitations of the one-loop calculation we have performed simulations
with
$\lambda\in[0,1.2]$ (see Fig.~\ref{fig:slacStrong}). The continuum extrapolations of Wilson type fermions
are only applicable up to $\lambda\le 0.7$ using lattice sizes of $\Ns\le 64$
due to the improvement problems. To cope with this,
we instead use SLAC fermions which allow for a much larger $\lambda$
range on the accessible lattice sizes.

For $\lambda>0.6$ the improved and unimproved model with SLAC fermions give
slightly different results on
a $45\times 45$ lattice. To check which model is closer to the continuum limit
additional simulations with $N=63\times 63$ at $\lambda=0.8$ have been performed
(Tab.~\ref{tab:slacContinuum}). The data unveil that the unimproved model
suffers from stronger finite $a$ effects. Therefore the 
correct continuum limit is reached for both SLAC models but the
\emph{improved SLAC model is closer to the continuum limit} on a finite lattice.


\section{Conclusions and outlook}
\begin{floatingtable}[r]
\begin{tabular}{lcc}\hline\hline
$\Ns$ & improved & unimproved \\ \hline
$45$ & $10.22(26)$ & $11.49(9)$ \\
$63$ & $10.54(15)$ & $10.70(19)$\\\hline\hline
\end{tabular}
\caption{\label{tab:slacContinuum}Fermionic masses for the SLAC derivative on
two different lattice sizes for $\lambda=0.8$.}
\end{floatingtable}
\myindent
We have performed a detailed analysis of the Nicolai improvement in the
Wess-Zumino
model. This improvement introduces new problems due to the sampling of unphysical
(high-momentum) states.  Additionally with high statistics bosonic and fermionic masses can \emph{not be
distinguished} for Wilson type fermions at finite lattice spacing in the weak to
intermediate coupling region for both the improved
and unimproved formulations. Even without improvement the \emph{correct
continuum limit} is reached. Therefore the term ``improvement'' is somewhat 
misleading. Only for SLAC fermions in the intermediate coupling
region the improved action is closer to the
continuum limit on finite lattices.

More detailed results on this model including the discussion of discrete
symmetries, the absence
of finite size effects and the effects of negative fermion determinants can be
found in \cite{Kastner:2008zc}.

In order to access the region of stronger couplings ($\lambda>1.5$) further
algorithmic improvements are necessary. With the help of the PHMC or RHMC algorithm and
improved solvers and preconditioners we are confident to obtain strong coupling
results in the near future.
The elaborate algorithms will then be used to explore the $\Ncal=1$ Wess-Zumino model in
two dimensions where a spontaneous supersymmetry breaking is expected and to
study supersymmetric $\mathbb{C}\mathrm{P}^N$ models in two dimensions.


\acknowledgments
\myindent
We thank S.~D\"urr for conversations about the determination of masses. Further
we thank P.~Gerhold and K.~Jansen for helpful
discussions concerning algorithmic details and Fourier acceleration. TK
acknowledges support by the Konrad-Adenauer-Stiftung
e.V., GB by the Evangelisches Studienwerk and CW by
the Studienstiftung des deutschen Volkes.
This work has been supported by the DFG grant Wi 777/8-2.



\begin{thebibliography}{9}
\small
\setlength{\baselineskip}{8pt}
\bibitem{Bergner:2007pu}
  G.~Bergner, T.~K\"astner, S.~Uhlmann and A.~Wipf,
  Annals Phys.\  {\bf 323}, 946 (2008).
\bibitem{Kastner:2007gz}
  T.~K\"astner, G.~Bergner, S.~Uhlmann, A.~Wipf and C.~Wozar,
  PoS {\bf LAT2007}, 265 (2007)
\bibitem{Bietenholz:1998qq}
  W.~Bietenholz,
  Mod.\ Phys.\ Lett.\  A {\bf 14}, 51 (1999).
\bibitem{Bergner:2008ws}
  G.~Bergner, F.~Bruckmann and J.~M.~Pawlowski,
  arXiv:0807.1110 [hep-lat].
\bibitem{Nicolai:1979nr}
  H.~Nicolai,
  Phys.\ Lett.\  B {\bf 89}, 341 (1980).
\bibitem{Beccaria:1998vi}
  M.~Beccaria, G.~Curci and E.~D'Ambrosio,
  Phys.\ Rev.\  D {\bf 58}, 065009 (1998).
\bibitem{Catterall:2001fr}
  S.~Catterall and S.~Karamov,
  Phys.\ Rev.\  D {\bf 65}, 094501 (2002).
\bibitem{Kastner:2008zc}
  T.~K\"astner, G.~Bergner, S.~Uhlmann, A.~Wipf and C.~Wozar,
  arXiv:0807.1905 [hep-lat].
\end{thebibliography}
\end{document}